# Chiral magnetic effect reveals the topology of gauge fields in heavy-ion collisions

**Dmitri E. Kharzeev[1,2] and Jinfeng Liao[3]**

1. Department of Physics and Astronomy, Stony Brook University, NY 11794-3800
2. Physics Department and RIKEN-BNL Research Center, Brookhaven National Laboratory, Upton, NY 11973-5000
3. Physics Department and Center for Exploration of Energy and Matter (CEEM), Indiana University Bloomington, IN 47405

Emails: dmitri.kharzeev@stonybrook.edu , liaoji@indiana.edu

*Abstract:*

The topological structure of vacuum is the cornerstone of non-Abelian gauge theories describing strong and electroweak interactions within the standard model of particle physics. However, transitions between different topological sectors of the vacuum (believed to be at the origin of the baryon asymmetry of the Universe) have never been observed directly. An experimental observation of such transitions in Quantum Chromodynamics (QCD) has become possible in heavy-ion collisions, where the chiral magnetic effect converts the chiral asymmetry (generated by topological transitions in hot QCD matter) into an electric current, under the presence of the magnetic field produced by the colliding ions. The Relativistic Heavy Ion Collider program on heavy-ion collisions such as the Zr-Zr and Ru-Ru isobars, thus has the potential to uncover the topological structure of vacuum in a laboratory experiment. This discovery would have far-reaching implications for the understanding of QCD, the origin of the baryon asymmetry in the present-day Universe, and for other areas, including condensed matter physics.

## *1. Introduction*

Quantum physics not only rules the microscopic world, but is also responsible for the defining properties, and quite possibly the very creation, of the entire Universe. We owe our existence to the baryon asymmetry generated shortly after the Big Bang in the newly created hot Universe. In 1967, Andrei Sakharov formulated three necessary ingredients of baryogenesis [1]: baryon number violation, violation of C(charge) and CP(charge-parity) symmetries and interactions out of thermal equilibrium. The latter condition was satisfied in the early Universe due to its rapid expansion, while quantum effects in the topologically non-trivial vacuum of the standard model (and its extensions) are crucial for satisfying the former two conditions. The existence of topologically distinct vacuum sectors is a direct consequence of the gauge symmetry that underlies the standard model. The compact nature of non-Abelian groups describing the electroweak and strong interactions allows for topologically non-trivial maps from the gauge space onto the Euclidean space-time [2], that in Minkowski space describe the tunnelling transitions [3] between the topological sectors of vacuum [4,5] characterized by different Chern-Simons numbers [6]. These quantum transitions (assisted by finite temperature [7]) constitute the crucial ingredient of the baryogenesis [8], and thus of our world.

The height of the potential barriers separating different vacuum sectors of the electroweak theory in our present world are set by the Higgs condensate. These barriers are most likely too high for collider experiments (although there are suggestions that this is not necessarily the case[9,10]), so the baryon number violation is probably not amenable to a direct experimental observation. Fortunately, the vacuum of Quantum Chromodynamics (QCD) possesses analogous topological vacuum sectors that are separated by much lower, and more easily penetrable, barriers. The experimental study of quantum transitions between these sectors can be performed in heavy-ion collisions that create a hot and rapidly expanding (thus out-of-equilibrium) QCD plasma. Heavy-ion experiments could thus recreate the conditions that existed in the early Universe a few microseconds after the Big Bang – and one of the most exciting possibilities that they provide is uncovering the structure of the vacuum responsible for baryogenesis. Of course, quantum transitions between the vacuum sectors of QCD do not generate a baryon number violation – but they do lead to an analogue violation of chirality. Away from equilibrium, this can result in a chiral asymmetry, the difference between the numbers of right- and left-handed quarks in the plasma.

Chirality (or handedness) is a key property of light fermions that is related to the sign of the projection of the fermion's spin onto its momentum – if this projection is positive, the fermion is right-handed, and if it is negative, the fermion is left-handed. In the case of anti-fermions, the positive (negative) projection corresponds to the left (right) chirality. In a classical theory, the chirality of massless fermions is conserved, but the quantum chiral anomaly [11,12] enables the transfer of chirality between fermions and gauge fields. This is what happens during quantum transitions between the vacuum sectors – in such transitions, the chirality of fermions is transferred to the chirality of the vacuum gauge fields, or vice versa. Since right- and left-handed fermions are related to each other by parity transformation P, the topological vacuum transitions thus induce violation of P and of combined charge C and parity CP symmetries. In direct analogy with the Sakharov criteria [1], the deviation from thermal equilibrium (driven by the rapid expansion of the produced matter) then enables a generation of net chirality in heavy ion collisions [13] – a "chirogenesis".

It is important to emphasize that QCD, in spite of the CP-violating vacuum transitions described above, does not possess a global CP violation. This means that micro-universes created in heavy-ion collisions will contain a net chirality of random sign – some will be produced preferentially right-handed, and some left-handed. It is not known what would happen to the sign of the baryon asymmetry in the Universe if it were repeatedly re-created – but heavy-ion collisions open the possibility to re-create billions of micro-universes, and to study the generation of the net chirality in this large statistical ensemble.

Although the opportunity to study an analogue of baryogenesis in a laboratory experiment is fascinating, the key problem in heavy-ion collisions is detecting the produced net chirality. After the created QCD matter expands and cools down, the chiral quarks become bound in massive hadrons, and the spontaneous breaking of chiral symmetry in the confined phase of QCD makes the net chirality unobservable. This looks like an insurmountable obstacle to the observation of "chirogenesis" – but a possible way of detecting it has been found [13].

The idea is based on the fact that quarks, apart from colour charge and chirality, possess an electric charge. Since the electric charge is strictly conserved, it is not affected by the conversion of quarks into hadrons – in other words, the electric charge of a blob of quark-gluon matter will be inherited by the produced hadrons. Electrically charged quarks also interact with external electromagnetic fields, and if chirality affects the flow of electric charge in an external magnetic field, this flow can be directly detected by measuring the electric charge of the final-state hadrons [13-16]. It appears that this possibility can indeed be realized in heavy-ion experiments (see Ref. [17] for a review), as we will describe below.

## 2. The chiral magnetic effect

Chirality plays a key role in QCD. In fact, the understanding of the role of chiral symmetry in strong interactions was key to formulating QCD. The theory possesses six 'flavors' of quarks (u (up), d (down), s (strange), c (charm), b (bottom), t (top)), and three of them (u, d, s) are light on the characteristic scale of strong interactions, $\Lambda_{QCD} \sim 200$ MeV. Because of this, a reasonable approximation of the physical world is an idealized theory with three massless quarks. The Lagrangian of this theory does not possess a single dimensionful parameter. Quantum effects due to the interactions with gluons, however, break down the scale symmetry of the theory, and induce the decrease of the coupling constant at short distances – this is the celebrated 'asymptotic freedom' [18, 19] of QCD. As a result, the dimensionful scale $\Lambda_{QCD}$ emerges ('dimensional transmutation') and therefore, a symmetry present in the Lagrangian is broken by quantum effects. This phenomenon is described as a quantum 'scale anomaly' [20,21]. A similar phenomenon affects the chiral property of massless quarks.

Because the massless quarks can be right- or left-handed, the theory possesses $U_R(3) \times U_L(3)$ chiral symmetry (U being the unitary group) describing two parallel worlds of right- and left- handed quarks. One may also express the corresponding symmetry group in a different way by introducing the vector $J_V = J_R + J_L$ and axial $J_A = J_R - J_L$ currents as superpositions of the currents of the right- and left-handed quarks, $J_R$ and $J_L$. In terms of vector and axial symmetries, $U_R(3) \times U_L(3) = SU_V(3) \times SU_A(3) \times U_V(1) \times U_A(1)$ (SU being the special unitary group). The vector parts of this symmetry manifest in the world of hadrons but there is no evidence for the axial part of the symmetry, which appears heavily broken – no mirror images (parity doublets) exist in the hadron spectrum. The ways in which the octet $SU_A(3)$ and the flavor-singlet symmetries $U_A(1)$ are broken in the physical world, however, are radically different. The octet part of the symmetry is broken spontaneously, akin to the way the rotational symmetry of a ferromagnet is broken below the Curie temperature. This breaking leads to the emergence of eight Goldstone bosons – pions, kaons, and the η mesons, and gives masses to other hadrons. This dynamically generated mass accounts for about 99% the mass of a proton or neutron, and thus for about 99% the mass of the visible matter in the Universe.

It is likely that this spontaneous breaking of the chiral symmetry occurs due to the effective interaction induced by instantons (see Ref. [22] for a review). The resulting Goldstone bosons play a key role in binding the nucleons in atomic nuclei. The flavor-singlet part of the symmetry $U_A(1)$ is broken by quantum interactions with gluons, similarly to the scale anomaly. This breaking is referred to as the chiral anomaly [11,12] discussed before. Because quarks possess both color and electric charges, the dynamics of the chiral anomaly can be explored not only with gluons, but also with photons. The photons can be directly observed in experiment; they can also be used to diagnose the behavior of quark-gluon matter through the chiral anomaly, which thus becomes a keyhole into the dynamics of quarks and gluons, and the way they exchange chirality.

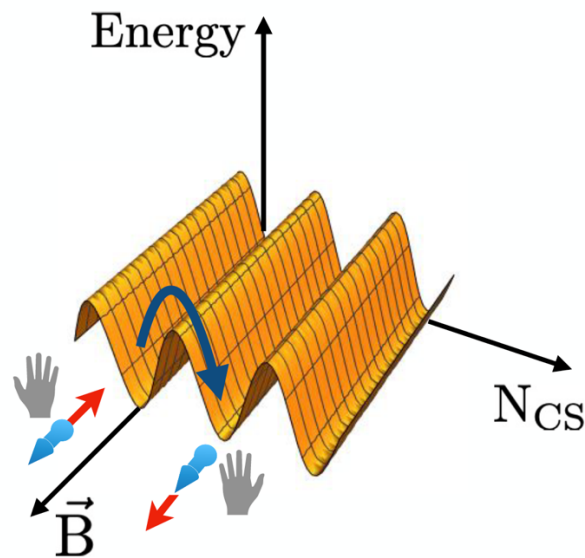

***Fig.1 An illustration of the mechanism that underlies the Chiral Magnetic Effect (CME) in quantum chromodynamics (QCD) matter** [13-16]. The QCD vacuum has a periodic structure, with minima corresponding to different Chern-Simons numbers that characterize the topology of color fields. An `instanton' [2] or `sphaleron' [7] transition between such energy-degenerate, but topologically distinct, vacuum sectors (shown by the curved dark blue arrow) is accompanied by the change of chirality of the chiral fermions. In an external magnetic field ($\vec{B}$) that pins down the direction of spin (blue arrows), the change of chirality has to be accompanied by the change in the direction of momentum (red arrows). If the numbers of left- and right-handed fermions are different, this results in an electric current along the direction of magnetic field – this is the CME.*

Indeed, consider a system of massless quarks in a strong magnetic field. Quantum charged particles in a magnetic field occupy a discrete set of Landau levels – closed orbits with different energies. For massless quarks, the lowest Landau level (LLL) has zero energy – the result of the cancellation between the positive kinetic energy of rotation and a negative Zeeman energy of the interaction of the quark's spin with a magnetic field. The direction of the quark's spin on the LLL is thus completely determined by its electric charge – the positive quarks (or antiquarks) will have their spins aligned along the magnetic field (see Fig. 1), and the negative ones – opposite to the magnetic field direction. The motion of quarks along the magnetic field is unrestricted – so, for the magnetic field pointing out of the plane of the Fig. 1, the quarks can move both out and into the plane, with an equal number of out- and into-the-plane movers in equilibrium. If a positive quark (with spin directed out of the plane) moves out of the plane, its projection of spin on momentum is positive, and we are dealing with a right-handed quark. If it moves into the plane, it is a left-handed quark. For negatively charged (anti)quarks with spin directed into the plane, the situation is the opposite – if they move out of the plane, they are left-handed, and if they move into the plane, they are right-handed. In equilibrium, the numbers of right- and left-handed quarks are equal, and there is thus no net electric current.

However, if there is an asymmetry between the densities of right- and left-handed quarks, this results into an electric current directed either along or against the direction of magnetic field, depending on the sign of asymmetry:

$$\vec{J} = \frac{e^2}{2\pi^2} \mu_5 \vec{B} \qquad (1)$$

– this is the Chiral Magnetic Effect (CME) [13-16] where $\vec{J}$ is the electric current and $\vec{B}$ is the magnetic field, the chiral chemical potential $\mu_5 = \mu_R - \mu_L$ characterizes the difference between the chemical potentials of the right- and left-handed quarks. For the effect to exist, the difference between the numbers of right- and left-handed quarks should not be conserved – in other words, the chirality of quarks should be transferable to the chirality of the gauge field configuration (described by the topological Chern-Simons number), and vice versa. This transfer is the essence of both the chiral anomaly and of the CME. In equilibrium, when the chiral chemical potential is fixed, the current in eq. (1) vanishes [23].

The simplest chiral configuration of electromagnetic fields is provided by parallel electric and magnetic fields. Indeed, the scalar product of electric and magnetic fields in the Coulomb gauge is proportional to the time derivative of the magnetic helicity (the Abelian Chern-Simons number) that measures the linking between the lines of magnetic flux [24, 25]. Parallel electric and magnetic fields thus pump chirality into the system, and due to chiral anomaly, the chirality of the gauge field can be transferred to the chirality of fermions, resulting in the CME. The reconnections of magnetic flux thus induce the chiral magnetic current [26].

In condensed matter physics, the analogs of chiral anomaly in the $^3$He – A superfluid phases have been discussed in Ref. [27]. 3D Dirac and Weyl semimetals possess emergent chiral quasiparticles (for reviews, see Refs. [28,29]) and thus offer the possibility to study the CME in a well-controlled setting. The CME was observed in experiments with 3D Dirac [30,31] and Weyl [32] semimetals where it leads to the longitudinal magnetoconductivity that grows quadratically with the strength of magnetic field. In very strong magnetic fields, this dependence is expected to become linear [16]. One also expects a new kind of quantum oscillations [33] (in addition to conventional Shubnikov – de Haas ones) resulting from the relation between the chiral chemical potential (that enters the formula for the chiral magnetic current) and the chiral density that is created due to the chiral anomaly. An interesting way to observe both the CME and the diffusion of the chiral charge through the material is a nonlocal transport measurement proposed in Ref. [34] and observed in Ref. [35]. A different way to use the chiral anomaly for the generation of the chiral magnetic current is to use circularly polarized light [36]. The correspondence between the currents induced by the chiral anomaly in different physical systems is shown in the Table 1 below.

Table 1: The currents induced by the chiral anomaly in different physical systems. The sources of chirality, the current carriers, the type of the induced anomalous current, and experimental signatures are indicated.

| System: | The Universe | Quark-Gluon Plasma | Dirac/Weyl semimetals | Superfluid $^3$He-A |
|---|---|---|---|---|
| Source of chirality: | Topological transitions in hot electroweak matter: sphalerons, … | Topological transitions in hot QCD matter: sphalerons, … | External electric and magnetic fields | Effective electric and magnetic fields induced by the time-dependent orbital angular momentum |
| Current carriers: | Quarks | Quarks | Electronic quasiparticles | Atoms in the superfluid |
| Type of the current: | Baryon | Electric | Electric | Linear momentum |
| Experimental signatures: | Baryon asymmetry of the Universe; Helical magnetic fields at intergalactic scales | Angular correlations of charged hadrons in relativistic heavy ion collisions | Negative longitudinal magnetoresistance; Non-local chiral transport; Chiral magnetic photocurrent | Dynamics of vortex motion |

The observation of CME in condensed matter systems establishes it as a calibrated tool that can be used to detect topological transitions in the QCD vacuum. In heavy-ion collisions, topological transitions are enhanced due to finite-temperature effects (`sphaleron' transitions analogous to thermal excitation processes). Because the produced matter has a finite size and expands rapidly, the system deviates from thermal equilibrium, therefore realizing the analogs of Sakharov criteria [1] for `chirogenesis'. Because the colliding heavy ions also creates an extremely strong magnetic field, the CME can occur and can be detected experimentally, as we will explain below.

## 3.. CME in the quark-gluon plasma

**From hadrons to quark-gluon plasma**

A non-perturbative scale $\Lambda_{\text{QCD}}$, on the order of 200 MeV, emerges due to the scale anomaly of QCD. Around this energy scale (or equivalently at distances around 1 fm = $10^{-15}$m), the QCD color force becomes very strong — about a hundred times stronger than the electromagnetic Coulomb force at the same distance. As a result, the quarks and gluons get confined within hadrons (for example, protons and neutrons), which thus become `cages` spanning about 1 fm across.

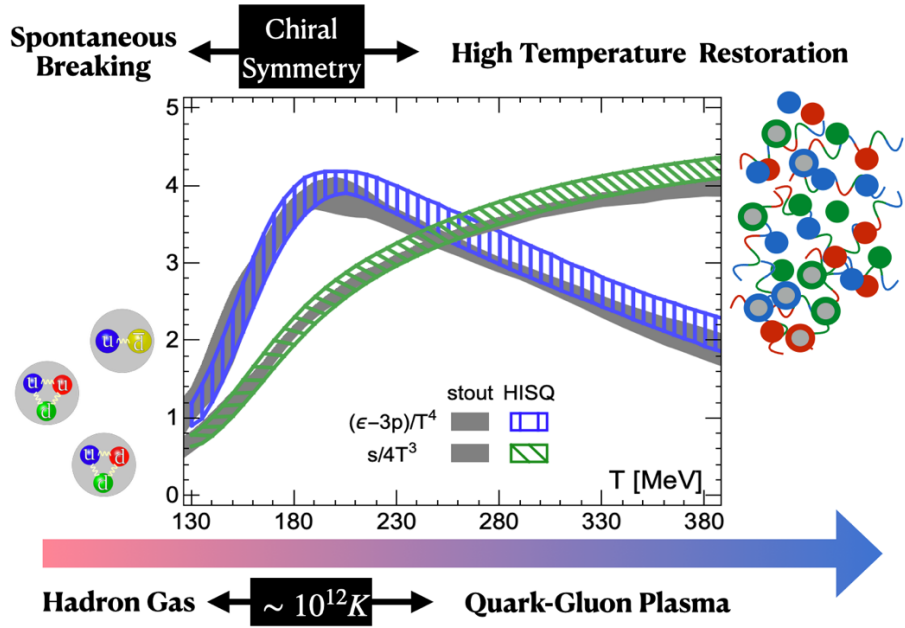

*Fig.2 Transition from hadron gas to quark-gluon plasma. At low temperature, the phase of nuclear matter is a hadron gas where quarks (u, d) and gluons are confined within hadrons such as protons, neutrons and pions (see left-hand-side). At high temperature, hadrons melt and quarks and gluons get liberated to form a new phase of nuclear matter called a quark-gluon plasma (QGP) (see right-hand-side). First-principle lattice simulations of Quantum Chromodynamics (QCD) [37,38], shown in the plot, find this deconfinement transition to be a rapid crossover at a temperature of about 155 MeV (on the order of $10^{12}$ Kelvin). The chiral symmetry of QCD is spontaneously broken in the low-temperature phase, where it generates most of the hadrons' masses. It gets restored around the temperature of deconfinement; in the high-temperature phase, the light-flavor quarks become (nearly) massless. Figure adapted with permission from Ref. [38].*

To penetrate the `iron curtain` set by $\Lambda_{\text{QCD}}$ and access quarks and gluons more directly, we need a high energy probe is needed. One way of doing this is to shoot a high energy electron or hadron at a proton or nuclear target. Another way is to heat up a chunk of QCD matter to a temperature exceeding $T \sim \Lambda_{\text{QCD}}$ which can be achieved by colliding relativistic heavy ions. It was envisioned in late 1970s that a new phase of matter — the quark-gluon plasma (QGP) — must emerge at some asymptotically high temperature scale. Powerful supercomputers have enabled the first-principle lattice QCD simulations that indeed confirm this idea and provide a fully quantitative understanding of QCD thermodynamics [37,38] (see Fig.2). The low temperature phase is a hadron gas composed of baryons (such as protons and neutrons) and mesons (such as pions and kaons), whereas the high temperature phase is a plasma of deconfined quarks and gluons. The deconfinement transition is a rapid crossover at temperature around 155 MeV (about $10^{12}$ Kelvin). Such temperatures were available a few microseconds after the birth of the Universe, so the QGP in fact is a `primordial' phase of matter. Today's high energy nuclear collision experiments at the Relativistic Heavy Ion Collider (RHIC) and the Large Hadron Collider (LHC) have recreated the QGP in the laboratory and have determined many key properties of this primordial matter (see Ref. [39] for a review).

The high temperature not only liberates the quarks from hadrons, but also leads to the restoration of chiral symmetry, much like the rotational symmetry of a ferromagnet is restored above the Curie point. This implies that the quarks become very light (and nearly massless for the up and down flavors) in the QGP phase. Because chirality is a well-defined property of massless quarks, the QGP thus provides the appropriate environment to explore the quantum chiral dynamics. Consider a heavy-ion collision in which a droplet of QGP is created in the overlapping `fireball' zone, (see Fig.3a). Due to the aforementioned topological fluctuations, the QGP randomly acquires a nonzero initial chiral charge. That is, on an event-by-event basis, a chiral QGP forms, with equal probabilities for its left-handed and right-handed guises. The fireball explodes rapidly outward with a radial flow velocity that could be as large as 60% or so of the speed of light. This rapid expansion, analogous to the expansion of the early Universe, allows the chiral charge to stay away from its zero equilibrium value (the `chirogenesis`). Therefore, it is expected that the quark-gluon plasma created in each single event of a heavy-ion collision is chiral.

Now the question is: can we observe the CME in the chiral QGP? For the CME to occur, one needs an external magnetic field, in addition to the chiral asymmetry. Fortunately, heavy-ion collisions not only create the highest man-made temperature, but also generate a pulse of very strong magnetic field [15]. The spatial distribution of magnetic field as computed in Ref. [40] is illustrated in Fig.3b. The initial nuclei in such a collision carry a large positive charge (for example atomic number Z=79 for the gold nucleus and Z=82 for the lead nucleus) and move at nearly the speed of light. As a result each of the colliding nuclei carries a huge magnetic field. In the overlap zone upon their collision, the magnetic fields $\vec{B}$ of the nuclei add coherently, and the resulting total magnetic field is orthogonal to the reaction plane, pointing in the out-of-plane ($y$) direction (see Fig.3a). Both simple estimates and quantitative simulations suggest that the magnitude of this magnetic field is on the order of $10^{15}$ Tesla, which is much stronger than the magnetic field on the surface of a magnetar (the previous record holder). For magnetic fields of this magnitude, electromagnetic interactions driven by this field are as strong as QCD interactions – in other words, the effects of magnetic field are no longer tiny corrections to strong interaction dynamics. The magnetic field, coupled with the chiral charge of the QGP, is thus expected to induce a CME current along the $\vec{B}$ field that transports charges across the reaction plane ($x$-$z$ plane in Fig.3a). This process results in a specific dipole pattern of the electric charge distribution corresponding to an out-of-plane charge separation, with experimentally measurable consequences that we will discuss next. The created magnetic field rapidly decays in time, making experimental observation more challenging. The time dependence of the in-medium magnetic field [41,42] is currently a major source of uncertainty in theoretical predictions.

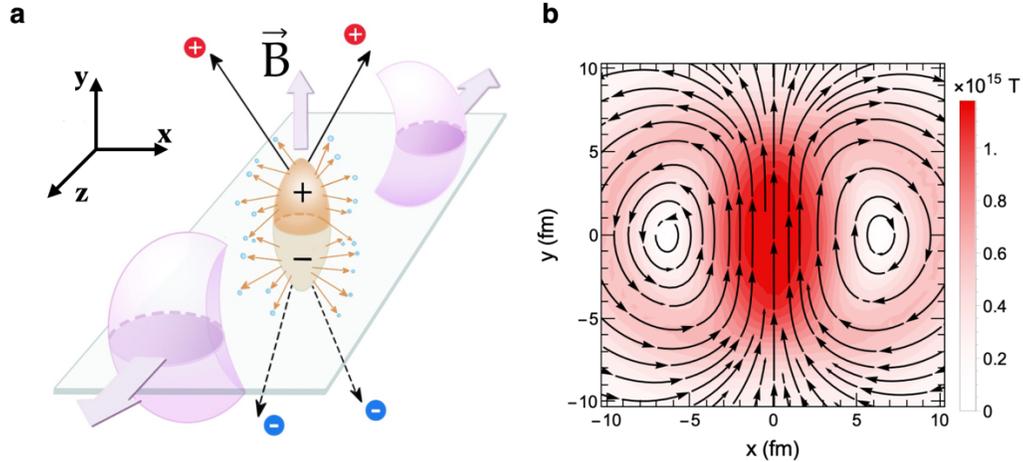

*Fig.3 Extremely strong magnetic field in a heavy-ion collision. a. A 3D illustration of a typical off-central heavy-ion collision is shown in the left panel. Two large nuclei initially travel at almost the speed of light down the beamline (z-axis) in opposite directions, with their centers displaced with respect to each other along the x-axis. Shortly after the impact, a hot quark-gluon plasma (QGP) forms in the overlap fireball zone (indicated as the ellipsoidal shape in the middle) where significant amount of kinetic energy from the initial nuclei is deposited and transformed into thermal energy. The side portions of the nuclei, with spectator nucleons not actively involved in the collision, continue to move along and part ways with the fireball. The spectator protons create a brief pulse of extremely strong magnetic field $\vec{B}$ that penetrates the QGP. This field points approximately along the out-of-plane direction (y-axis) that is perpendicular to the reaction plane spanned by x-z axes [40]. b. Quantitative simulation results for the distribution of magnetic field direction (indicated by small arrows) and strength (indicated by color scheme) on the x-y plane for gold-gold (AuAu) collisions at Relativistic Heavy Ion Collider (RHIC). In the center of the created QGP, the $\vec{B}$ field reaches an extreme magnitude, on the order of $10^{15}$ Tesla or more, which is the strongest magnetic field known in today's Universe. New effects arising from such extreme magnetic field have been explored both experimentally and theoretically [43-48]. Panel a is adapted with permission from Ref. [69]; panel b is adapted with permission from Ref. [39].*

The colliding system in a (typically non-central) heavy-ion collision possesses a large angular momentum arising from the fact that the two incident nuclei carry large and opposite momenta with a nonzero displacement between their respective centers of mass. Simulations show that the fireball carries an angular momentum on the order of $10^{4\sim5}\hbar$ (that is hundred thousand times the angular momentum of a proton) at RHIC (and even larger at LHC) over a volume just about hundred times that of a proton. A fraction of this angular momentum stays within the QGP and leads to very rich fluid vorticity patterns. Like the magnetic field, such vorticity is at extreme, as large as ~$10^{22}\ s^{-1}$ at RHIC. This is much faster than any other known fluid rotation, and thus the fireball in heavy-ion collisions is also considered "the most vortical fluid"[49]. Vorticity leads to a number of interesting effects, including the Chiral Vortical Effect [14,50-54] that is similar to the CME. In 2017, the STAR Collaboration at RHIC discovered the global spin polarization of produced hadrons, induced by the fluid vorticity at the subatomic scale[49]. This result has attracted a lot of interest and led to a number of new developments in the quantum theory of rotating fluids. Some of these effects were also observed experimentally in condensed matter systems, see for example Refs [27,55,56].

**Hunting for CME in heavy ion collisions**

Let us now focus on the search for CME signals in heavy-ion collisions. The CME predicts a separation of positive and negative charges in the QGP fireball along the axis of magnetic field. To illustrate how this happens in a real-world collision event, the charge separation in the *x-y* plane transverse to the beam axis *z* is shown in Fig.4a. Due to fluctuations in the positions of the nucleons inside the incident nuclei, the shape of the created fireball does not really look almond–shaped as depicted in simple cartoons (as in Fig.3a), but instead looks lumpy and distorted. Nevertheless, one could experimentally identify a `principal axis' (indicated by $\Psi_2$ in Fig.4a) for the dominant elliptic shape in each collision event. This axis and the beam axis together define the event plane. Similarly, due to fluctuations, the direction of magnetic field $\vec{B}$ is close to, but is not perfectly aligned along the *y*-axis. Such azimuthal fluctuations and the correlations between the $\vec{B}$ and $\Psi_2$ directions can be quantified: on average, they are perpendicular to each other, but with a sizable spread [40]. As we shall see, these fluctuations are important to take into account in experimental measurements. In practice, experimentalists identify $\Psi_2$ in every event and infer the $\vec{B}$ direction with the help of simulations.

The CME induces a charge separation along $\vec{B}$ field for the quarks in the QGP, for example with more positive quarks on one side of the fireball and more negative quarks on the other side. Upon hadronization (the conversion process from QGP into a hadron gas) at the late time of a collision, the charge separation pattern is inherited by the resulting charged hadrons. This geometric pattern, coupled with a strong radial flow (driven by the pressure gradients in QGP) leads to a specific signal: for a given event, the positive hadrons are preferably emitted along $\vec{B}$, whereas the negative hadrons are emitted in the opposite direction, or vice versa. To make an analogy, this is measuring an electric dipole moment (EDM) of the entire QGP.

Depending on the chirality of the QGP fireball, the CME current is either parallel or anti-parallel to $\vec{B}$. That means that the correspondence between positive/negative charges and north/south emission angles would flip from event to event with equal probability for the two configurations. Simply measuring the anticipated dipole pattern of the charged hadrons by averaging over many collision events would give a null result. This is not surprising: as previously mentioned, QCD does not break parity globally and thus on average the QGP should have zero EDM. All we can hope for is to measure the variance ( that is the `square') of this event-by-event charge separation pattern.

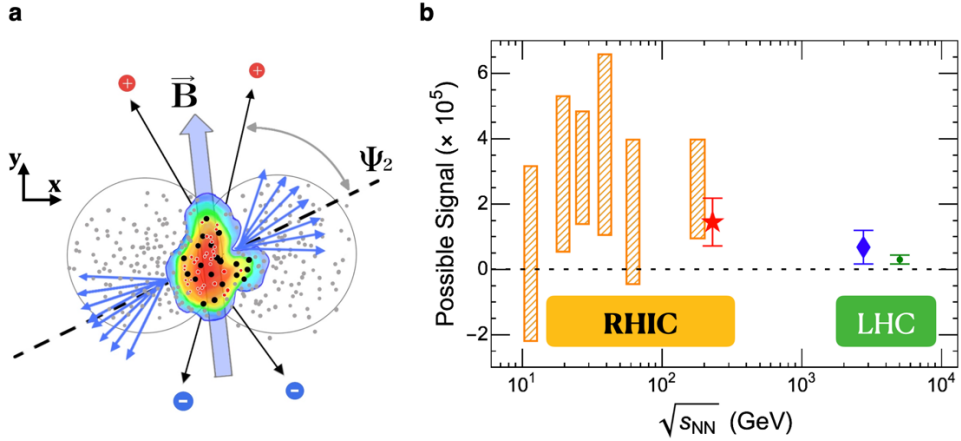

*Fig.4 Experimental measurements of the Chiral Magnetic Effect (CME). a. An illustration of the hadron angular correlation pattern in the (x-y) plane transverse to the beam axis z in a heavy-ion collision. The CME induces an asymmetry in the emission of positive and negative hadrons along the axis of magnetic field $\vec{B}$ that is approximately orthogonal to the event plane angle $\Psi_2$. b. Possible CME signal, as extracted from the measured charge asymmetry `$\gamma$ −correlator' (see text for the definition), is shown for heavy-ion collisions at Relativistic Heavy Ion Collider (RHIC) and Large Hadron Collider (LHC) over a broad range of the center-of-mass energies per nucleon pair $\sqrt{s_{NN}}$. The data in panel b was compiled by the authors based on the experimental results published in Refs. [65,66,71,78,79], comprising different centrality ranges and with various background assumptions. Panel a is reproduced with permission from Ref. [88].*

Experimentalists found a clever way of doing just this, by measuring the angular correlations between charged hadrons. Despite the dipole orientation being either along or against the direction of magnetic field, the emission pattern (see Fig.4a) is such that along the axis perpendicular to the event plane $\Psi_2$, the strong radial flow pushes extra positive charges to move together in one direction, while the extra negative charges move together in the opposite direction. As a result, two same-sign (SS) hadrons tend to be produced side-by-side whereas two opposite-sign (OS) hadrons tend to be produced back-to-back. These charge-dependent two-hadron correlation patterns remain the same despite flipping the orientation of the CME-induced dipole in the fireball. The difference between the angular correlations of SS and OS pairs can thus be a signal of the CME, as first proposed in Ref. [57]. Such charge asymmetry correlation measurements can be done through a number of carefully crafted observables, for example the so-called $\gamma$ − and $\delta$ −correlators [57] (and their variants [58-62]), event-by-event shape analysis [63-66], the R-correlator [67] and charged balance function [68]. Extensive experimental efforts have been carried out at both RHIC and LHC over the past decade to measure these observables for collisions spanning a wide range of center-of-mass energies [64-66,69-77]. Although they do demonstrate sensitivity to the CME signal, it turns out that they are unfortunately susceptible to a number of background correlations, see further discussion in Refs [17,39,58,78,79,80]. For example, many hadron resonances emerging from the collision decay into secondary hadrons which often contain

OS pairs and rarely contain SS pairs. Furthermore, these decay products are correlated in their momentum directions as they inherit a fraction of the parent resonance's momentum that is affected by hydrodynamical flow of the QGP. As a result, the resonance decays give a background (that is, non-CME) contribution to the charge asymmetry correlation. A few other identified background contributions exist as well.

To make it worse, a detailed analysis shows that in the overall charge asymmetry correlation, non-CME backgrounds dominate, and the CME signal is relatively weak. The hunt for the CME thus faces a signal isolation challenge similar to the search for the Higgs boson in hadronic collisions or the search for tiny temperature fluctuations in the cosmic microwave background radiation. Luckily the background contribution and the CME signal are driven by different features of a collision event, with the former controlled by the fireball's elliptic anisotropy while the latter controlled by the magnetic field. Based on this difference, a number of strategies were developed to separate the signal and background in the measured correlation. Such analyses indeed confirm that the CME signal is relatively small, likely accounting for not more than 10% of the total correlation. A compilation of the current extractions of possible CME signal for collisions at various beam energies is presented in Fig.4b. Caution must be taken as these extractions are often subject to model assumptions and/or poorly controlled systematic uncertainties. Nevertheless, these experimental results, although far from being conclusive, are strongly suggestive of a detectable CME signal, especially in the RHIC energy region.

There exist other interesting effects that could be measurable. A notable example is the so-called Chiral Magnetic Wave (CMW), which is a manifestation of CME through a gapless collective mode of the chiral density wave [81]. The CMW is predicted to induce a quadrupole pattern of charge distribution in the QGP fireball [82,83] which can be measured via charge asymmetry in the quadrupole component of the hadrons' angular distribution (the `elliptic flow' coefficient). Intriguing evidence for this effect was also reported at RHIC and LHC [84,85].

To summarize, the challenge in the search for CME is the isolation of the weak signal embedded in strong backgrounds. Different approaches were developed to overcome this difficulty, the majority focusing on ways to separate the backgrounds from the signal in the measured correlations, based on their differing dependence on collision environments (for example, beam energy, system size, event shape) or on observable kinematics (for example, the invariant mass of hadron pairs). This method, applied to the large amount of available data, helps achieving a systematic quantification of both signal and backgrounds, but suffers from uncertainties due to various model assumptions about the origin of the backgrounds. There are also attempts to devise observables arguably insensitive to backgrounds which look very promising, but need a more in-depth scrutiny to be fully validated. CMW measurements can provide alternative evidence for the CME, but they require considerably more statistics and further examination of potential background contributions. Apart from all these, it is highly desirable to have a `clean', essentially model-independent method that can decisively reveal the existence of CME even if the backgrounds are not completely understood. The isobar collision experiment provides such an opportunity, as we discuss next.

## *4. The isobar collision experiment*

Observing CME in heavy-ion collisions is of paramount importance, but the corresponding measurements face a severe challenge due to a strong background contamination. This situation calls for a new experimental approach, and for a quantitative characterization of both signal and background.

On the experimental front, the idea to collide and contrast a pair of isobar nuclei emerged, matured and was realized [86,87,88]. A dedicated isobar collision run was successfully carried out at RHIC in 2018. This experiment can be decisive in the search for the CME. In this experiment, one compares the outcome of Ruthenium-Ruthenium (RuRu) and Zirconium-Zirconium (ZrZr) collisions. Ru and Zr are so-called isobar nuclei, with the same number of nucleons (A=96 for both), but different number of protons (Z=44 for Ru and Z=40 for Zr). This means that Ru and Zr nuclei have roughly the same size and mass so that the bulk fireball created in RuRu and ZrZr collisions would be nearly identical. However, the magnetic field strength, which is proportional to the nuclear charge Z, would differ by about 10% between the isobar pairs. This expectation is indeed confirmed by quantitative simulation results in Fig.5a, where the relative difference between the two colliding systems is found to be negligible for the fireball geometric anisotropy (shown as $\Delta\langle\epsilon_2\rangle$) and to be around (10~20)% for the square of magnetic field (shown as $\Delta\langle B_{sq}\rangle$). The contrast between RuRu and ZrZr thus offers a unique opportunity for the CME search. This is because the CME signal is driven by the magnetic field whereas the background is controlled by the fireball elliptic anisotropy. One would then expect that for the charge asymmetry correlation measurement, there should be an equal amount of background contributions in both RuRu and ZrZr systems whereas a different amount of CME signal contributions, as illustrated in Fig.5b. So a detectable variation of the charge asymmetry correlation from RuRu to ZrZr collisions should only arise from CME and can thus serve as its unambiguous signature.

The precise amount of difference in the charge asymmetry correlation between RuRu and ZrZr would depend upon the signal-to-background ratio. If the background level is too high, the isobar contrast might become too small to be detected. Fortunately, during the 2018 isobar run the STAR Collaboration collected about 3 billion collision events for each system, providing a very strong differentiating power enabled by high statistics. The STAR Collaboration projection for the experimentally measurable isobar difference level as a function of background contribution is shown in Fig 5c. As a benchmark, if the background contribution is no more than 86%, then the isobar measurement would discover the CME with $5\sigma$ significance.

On the theory front, significant progress was made in describing chiral transport in and out of equilibrium [89-96], and in phenomenological applications for isobar collisions [97-105]. In particular, a state-of-the-art simulation tool, known as EBE-AVFD (event-by-event anomalous-viscous fluid dynamics) [97-99], has been developed in the last couple of years. This framework allows a quantitative characterization of both the CME signal and the background contribution in realistic heavy ion collision environment. The tool is now widely used for studying CME-related observables. A set of predictions for isobar collision experiment has been made from EBE-AVFD [99] (see Fig 5d). These predictions provide valuable inputs for the ongoing isobar analysis and demonstrate specific features of the CME signal that will soon be tested in experiment [106].

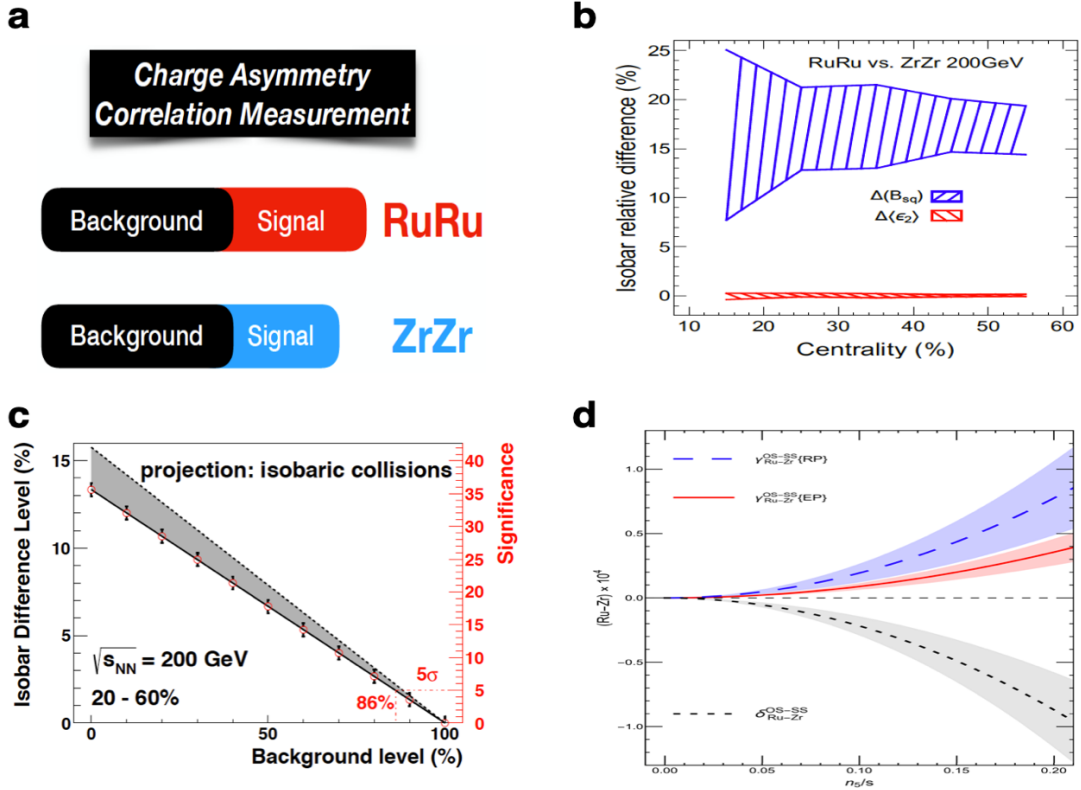

*Fig.5 Chiral Magnetic Effect (CME) in isobar collisions*. **a**. *Simulations of the RuRu and ZrZr collisions [99] show a negligible relative difference for the fireball geometric anisotropy $\Delta\langle\epsilon_2\rangle$ and a sizable difference for the square of magnetic field $\Delta\langle B_{sq}\rangle$. The bands indicate uncertainty arising from fluctuations and ambiguity in the initial nuclear geometry [100-102].* **b**. *The background contribution to the charge asymmetry correlation measurement is controlled by bulk geometric anisotropy and thus expected to be identical between RuRu an ZrZr systems. However, the CME signal is driven by magnetic field and should be different.* **c**. *The projection for the isobar difference level in experimental observables estimated as a function of background contribution [87]. The grey band indicates uncertainty from calculated geometric anisotropy and magnetic field for RuRu and ZrZr systems while the error bars indicate expected statistical uncertainty, assuming 3 billion collision events for each colliding system recorded by STAR in 2018. $\sqrt{s_{NN}}$ is the center-of-mass energy per nucleon pair.* **d.** *Quantitative predictions from event-by-event anomalous-viscous fluid dynamics (EBE-AVFD) simulations [99] for the difference in γ- and δ-correlators between the isobar pairs. Panels b, c are reproduced with permission from Ref. [88]; panel d is reproduced with permission from Ref [87].*

## 5. Perspectives

If the isobar experiment at RHIC establishes the existence of the Chiral Magnetic Effect driven by topological transitions in the quark-gluon plasma, this discovery will open new research pathways in nuclear physics and beyond. In nuclear physics, one will be able to study topological transitions in baryon-rich matter, and possibly in the vicinity of the critical point on the QCD phase diagram, using the beam energy scan that is underway at RHIC. These measurements will also allow to extract the rate of topological transitions that is poorly known theoretically – this may offer unique insights into the dynamics of baryogenesis.. In condensed matter physics, the study of Chiral Magnetic Effect and the underlying topology has already begun but has been limited so far to transport measurements. In the near future, these studies will be extended to chiral magnetic currents driven by light and strain. Among the possible future applications of Chiral Magnetic Effect are the ``chiral qubits" and quantum sensors.

**Acknowledgements**
This work is partly supported by the U.S. Department of Energy, Office of Nuclear Physics, within the framework of the Beam Energy Scan Theory (BEST) Topical Collaboration. The authors also acknowledge support by the U.S. Department of Energy, Office of Nuclear Physics Contracts No. DE-FG-88ER40388 and No. DE-SC0012704 (DK), and by NSF Grant No. PHY-1913729 (JL). We are grateful to B. Liao, S. Mukherjee, S. Shi, P. Tribedy, G. Wang, and H. Zhang for valuable help.